\documentclass{article}

\usepackage[utf8]{inputenc}
\usepackage{epsfig}
\usepackage{amssymb,amsmath}
\usepackage{latexsym}
\usepackage{gensymb}

\usepackage{float}
\usepackage{graphicx}
\usepackage{caption}
\usepackage{multirow}

\title{Comparison between Laplace-Lagrange Secular Theory and Numerical Simulation}
\author{Barbara C.B. Camargo\\
 {\small Sao Paulo State University (UNESP), Sao Paulo, Brazil}\\
 \and Othon C. Winter\\
 {\small Sao Paulo State University (UNESP), Sao Paulo, Brazil}\\
 \and Dietmar W. Foryta \\
{\small Federal University of Parana (UFPR), Physics Department, Parana, Brazil}}

\date{version: \today}

\begin{document}

\maketitle

\begin{abstract}
\noindent The large increase in exoplanet discoveries in the last two decades
  showed a variety of systems whose stability is not clear.  In this
  work we chose the $\upsilon$ Andromedae system as the basis of our
  studies in dynamical stability. This system has a range of possible
  masses, as a result of detection by radial velocity method, so we
  adopted a range of masses for the planets $c$ and $d$ and applied
  the secular theory. We also performed a numerical integration of the
  3-body problem for the system over a time span of 30 thousand years.
  The results exposed similarities between the secular perturbation
  theory and the numerical integration, as well as the limits where
  the secular theory did not present good results. The analysis of the
  results provided hints for the maximum values of masses and
  eccentricities for stable planetary systems similar to $\upsilon$
  Andromedae.
\end{abstract}

\section{Introduction}\label{sec:1}

In the last 20 years there has been a large increase in exoplanet
discoveries. We know the existence of about three thousand planets and
more than two thousand candidates, which have varied features (Han et
al., 2014). However, the dynamical stability for many of these planets
is not clear yet.

The $\upsilon$ Andromedae was one of the first exoplanetary systems
discovered (Butler et al., 1997) using the radial velocity technique,
but this method is subject to uncertainty regarding the relative
values of the masses and inclinations of the system's members.

Using the velocity measurements of the Hobby-Eberly telescope combined
with the astrometric data obtained by the Hubble Space Telescope,
McArthur et al. (2010) refined the orbital parameters and determined
the orbital inclinations for the planets $b$, $c$ and $d$ (see table
1) as well as the longitudes of pericenter and ascending nodes of the
planets. Using these inclinations and the mass of the star as $1.31$
solar masses, they determined the mass of planets $c$ as $14.5 M_J$
and $d$ as $10.2 M_J$ , where $M_J$ is the mass of Jupiter.

On the other hand, Curiel et al. (2011) adopted coplanar orbits and
minimum masses for the planets $c$ and $d$. The masses are calculated
to be $1.9 M_J$ and $4.1 M_J$ for the planets $c$ and $d$,
respectively.  Curiel et al. (2011), looking for planetary systems
with large residues after subtracting the 3-body models, found that
the $\upsilon$ Andromedae residues show a radial velocity which
suggests the presence of an additional long-period orbit in the
system, so the presence of a fourth planet, $e$, is possible in this
planetary system.

We can notice that these two models have a large discrepancy between
the estimated planetary masses. In order to help discriminate the more
acceptable mass values for this system, we studied the stability of
them according to different masses considered for planets $c$ and $d$.
Planet $b$ was not examined due to its proximity to the star. The
planet $e$ is not predicted in the study of McArthur et al. (2010), so
we also did not include it in this study.

One approach to study the stability of a pair of planets (planets $c$
and $d$) is to first check the evolution of the orbital eccentricity.
The theory of secular perturbation can provide such information as a
first approximation. The limits of the validity of the secular theory
can be found through full numerical integrations.  In order to
identify the actual unstable trajectories we performed numerical
integration of the 3-body problem for star, planet $c$ and planet
$d$. This study can provide a hint on the upper limits to the values
of planetary masses that can keep the system in a stable
configuration.

In Section \ref{sec:2}, a brief theoretical introduction of the
secular perturbation theory for the three-body problem is
presented. Section \ref{sec:3} shows the results obtained by
simulating the secular theory and the numerical integration of the
full equations of motion. In Section \ref{sec:4}, we discussed the
validity of the secular perturbation for $\upsilon$ Andromedae system.
In Section \ref{sec:5}, we discuss the implications of the results in
terms of the limiting values of the planetary masses.

\section{Method}\label{sec:2}

In this work we will focus on the stability of the $\upsilon$
Andromedae planetary system to estimate a limit for the masses of the
planets $c$ and $d$. In order to do that, we will track the evolution
of the eccentricities of these planets starting from the orbits
predicted by Curiel et al. (2011), which assumed planar orbits
different from the work of McArthur et al. (2010), which assumes
inclinations on the planets orbits.

So, as a first approximation, the methodology in this work will
consider the secular theory presented in Murray \& Dermott (1999),
where two bodies $m_1$ and $m_2$ interact with each other while
orbiting a central mass $M_c$ ($m_1, m_2 \ll M_c$).

Thus, assuming planar orbits and small initial eccentricities, we can
derive the secular evolution based on the Laplace-Lagrange theory,
expanded to second order in eccentricities, this results in a
disturbing function given by Murray \& Dermott (1999):

\begin{equation}
    R_j = n_ja_j^2 \left[\frac{1}{2} A_{jj}e_j^2 + A_{jk}e_1e_2\cos( \varpi_1-\varpi_2 ) \right],
\end{equation}
where $j=1,2$, $k=2,1$ $(j\neq k)$, and 
\begin{align}
  A_{jj} = +n_j \frac{1}{4} \frac{m_k}{m_c + m_j}\alpha_{12}\bar{\alpha}_{12}b_{3/2}^{(1)}(\alpha_{12}),\\
  A_{jk} = -n_j \frac{1}{4} \frac{m_k}{m_c + m_j}\alpha_{12}\bar{\alpha}_{12}b_{3/2}^{(2)}(\alpha_{12}),
\end{align}
where $\bar{\alpha}_{12}=\alpha_{12}$ if $j=1$ (external perturbation)
and $\bar{\alpha}_{12}=1$ if $j=2$ (internal perturbation), the
$b^{(i)}$ correspond to Laplace's coefficients, $\varpi$ denotes the
longitude of pericenter, $a$ is the semi-major axis, $e$ is the
eccentricity and the reference orbit has osculating elements
associated with $n^2 a^3 = G M_C$, where $n$ is the mean motion.

Therefore, to avoid singularities that are inherent in the equations
for small values of eccentricity, new variables are introduced, $h_j =
e_j\sin \varpi_j$ and $k_j = e_j\cos \varpi_j$. Using the new
variables, the differential equations can be expressed as
\begin{equation}
  \begin{split}
    \dot h_j = {+} \frac {1}{n_ja_j^2} ~\frac {\partial R_j}{\partial k_j} ,\\
    \dot k_j = {-} \frac {1}{n_ja_j^2} ~\frac {\partial R_j}{\partial h_j} , 
  \end{split}
\end{equation}

Then, the solutions for equations (4) are given by 
\begin{equation}
  \begin{split}
    h_j = \sum_{i=1}^{2}e_{ij}\sin (g_it + \beta_i) ,\\
    k_j = \sum_{i=1}^{2}e_{ij}\cos (g_it + \beta_i) , 
  \end{split}
\end{equation}
where the frequencies $g_i$ ($i=1,2$) are eigenvalues and $e_{ij}$ are
the components of the eigenvectors related to the matrix corresponding
to the elements formed by equations (2) and (3).  The phases $\beta_i$
are determined by the initial conditions.  The solutions described in
equations (5) are the classic secular solution of Laplace-Lagrange
secular problem (Murray \& Dermott, 1999).

 In the work of Gomes et al. (2006) is proposed an approach for
  systems with large values of eccentricities because the secular
  solution given by equations (5) would be only valid for small values of
  eccentricities.  They propose that for large values of eccentricity
  becomes necessary to incorporate the averaged effect of an eccentric
  orbit upon the motion of the perturbed planet. The averaged effect
  is computed assuming that the perturber is on a circular orbit
  of radius $b$, where
\begin{equation}
b = a\sqrt{(1 - e^2)}, 
\end{equation}
and $a$, $e$ are semi-major axis and eccentricity of the perturber’s
real orbit. (Gomes et al. , 2006). 

\begin{figure}[!t]
  \noindent
  \begin{tabular}{lr}
    \includegraphics[width=0.48\textwidth]{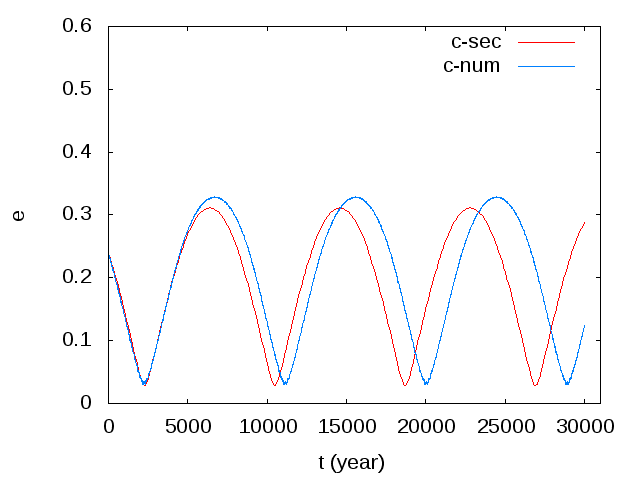}& 
    \includegraphics[width=0.48\textwidth]{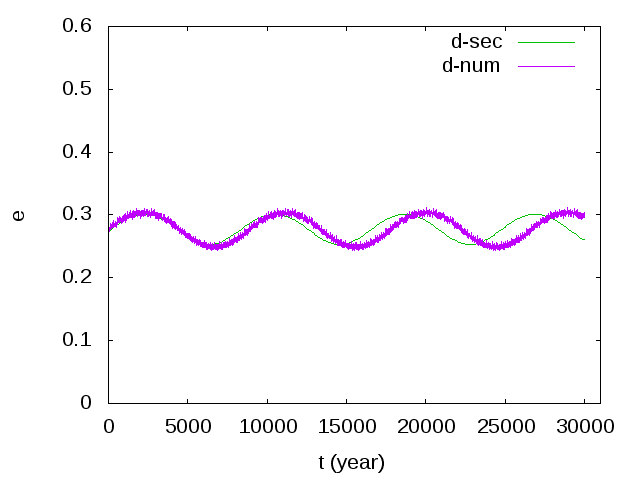}\\
  \end{tabular}
  \caption{Evolution of eccentricity for the case of $m_c = 2 M_J$ and
    $m_d = 4 M_J$, using the secular perturbation, and the full numerical
    integration. On the left plot, we present the evolution of
    eccentricity for planet $c$.  On the right plot, we present the evolution of
    eccentricity for planet $d$.}
  \label{fig:1}
\end{figure}

In figure \ref{fig:1}, we can see a comparison between
the secular theory  and the full numerical integration.  We used secular theory 
solution, equations (5), to analyse the eccentricity evolution 
for the case of $m_c = 2 M_J$ and $m_d = 4 M_J$ using the 
parameters of Curiel et al.(2011) (Table 1). We performed
numerical simulations of the planar case of a 3-body problem for 
30 thousand years, using the {\it Mercury} package (Chambers, 1999).
Based on these results seems that secular theory without correction is
enough for the dynamical system studied in the present work. Therefore, 
we will not use the correction suggested by Gomes et al. (2006).

\section{Numerical Simulations}\label{sec:3}

\begin{table}
 \caption{Orbital Elements for $\upsilon$ Andromedae planets  from 
    McArthur et al. (2010) and Curiel et al. (2011).}
  \label{tab:1}
  \begin{tabular}{l l l l}
    \hline\noalign{\smallskip}
                                      &   & McArthur et al. (2010) & Curiel et al. (2011) \\
    \hline\noalign{\smallskip}
    $M_*$ (M$_\odot$)                 &   & 1.31                   & 1.30                 \\
    \hline\noalign{\smallskip}
    \multirow{4}*{$M_p$ ($M_J$)}      & b & 5.9                    & 0.6876               \\
                                      & c & 14.57                  & 1.981                \\
                                      & d & 10.19                  & 4.132                \\
                                      & e & -                      & 1.059                \\
    \hline\noalign{\smallskip}
    \multirow{4}*{$a$ (au)}           & b & 0.059                  & 0.0592               \\
                                      & c & 0.861                  & 0.8277               \\
                                      & d & 2.703                  & 2.5133               \\
                                      & e & -                      & 5.2455               \\
    \hline\noalign{\smallskip}
    \multirow{4}*{$I$ (\degree)}      & b & 6.9                    & -                    \\
                                      & c & 16.7                   & -                    \\
                                      & d & 13.5                   & -                    \\
                                      & e & -                      & -                    \\
    \hline\noalign{\smallskip}
    \multirow{4}*{$e$}                & b & 0.010                  & 0.0215               \\
                                      & c & 0.239                  & 0.2596               \\
                                      & d & 0.274                  & 0.2987               \\
                                      & e & -                      & 0.0053               \\
    \hline\noalign{\smallskip}     
    \multirow{4}*{$\Omega$ (\degree)} & b & 45.5                   & -                    \\
                                      & c & 295.5                  & -                    \\
                                      & d & 115.0                  & -                    \\
                                      & e & -                      & -                    \\
    \hline\noalign{\smallskip}
    \multirow{4}*{$\omega$ (\degree)} & b & 41.4                   & 324.9                \\
                                      & c & 290.0                  & 241.7                \\
                                      & d & 240.8                  & 258.8                \\
                                      & e & -                      & 367.3                \\
    \noalign{\smallskip}\hline
  \end{tabular}
\end{table}

To verify the orbital stability of the system for different mass
values for planets $c$ and $d$, we studied the temporal evolution of
the eccentricity of each planet. In this study, we adopted the initial
values of $a$, $e$ and $\omega$ shown in Table 1 from Curiel et
al. (2011).

In the secular theory is assumed that $r_1<r_2$, where $r_1=r_c$ and
$r_2 =r_d$ are the distances from the central body for planets $c$ and
$d$, respectively.  In other words, the orbits of the two planets
cannot cross. This limitation can be characterized considering the
situation where the distance from the apocenter of the inner body
(planet $c$) is equal to the distance from the pericenter of the outer
body (planet $d$), i.e.:
\begin{equation}
  \begin{split}
    a_c  (1 + e_c) = a_d (1 - e_d) ,
  \end{split}
\end{equation}
where $a_c$, $a_d$ are the semi-major axes of the planets $c$ and $d$
while $e_c$, $e_d$ are the eccentricities of the planets $c$ and $d$,
respectively.

\begin{figure}[!t]
  \centering
  \includegraphics[width=0.8\textwidth]{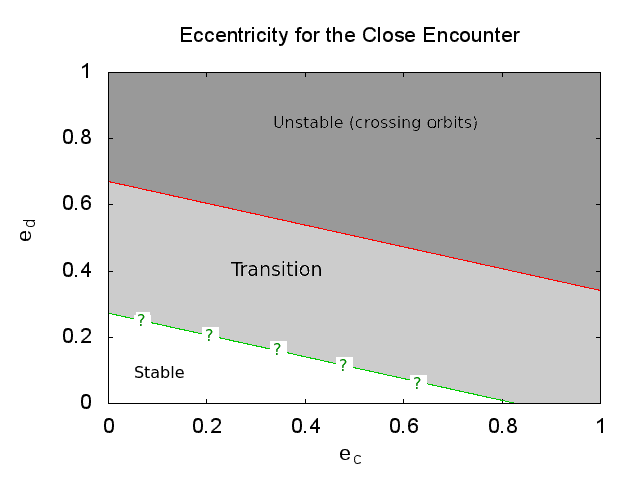}\\
  \caption{Relation between the critical eccentricities of the planets
    $c$ and $d$, according to eq. (7). The dark gray region indicates
    an instability area for the orbits of planets $c$ and $d$. Between
    the unstable and stable region (white area), it is located the
    transition region which takes into account the gravitational
    interactions between two masses even not necessarily crossing their
    orbits.}
  \label{fig:2}
\end{figure}

Since the secular perturbation does not affect the values of the
semi-major axis, we have $a_c$ and $a_d$ remaining constants. In this
way, we can get the values of the critical eccentricities that
indicate when the secular theory is certainly not valid and
instability occurs in the system. In Figure \ref{fig:2}, we show the
relation between the eccentricities of planets $c$ and $d$, given by
eq.(7). The dark gray region indicates the unstable orbits induced by
the close encounter between the planets.

For the value of $e_c = 0.2596$ (Curiel et al, 2011), we find the
critical eccentricity of planet $d$ as $0.605$. For the value of $e_d
= 0.2987$ (Curiel et al, 2011), we find that the orbit of planet $c$
should be hyperbolic.This result only takes configurations where the
planets'orbits cross.

In fact, even without crossing orbits, there may appear instability
due to the gravitational interactions between the two planets. So,
there is a minimal distance between the two orbits that can produce
instability in the planetary orbits, creating a region indicated by
``Transition'' in Figure \ref{fig:2}. The location and size of such
region depends on the masses of the planets involved.  Therefore, in
Figure \ref{fig:2} we divided the $e_c \times e_d$ plot in three
regions: one region that is unstable due to the crossing of the
orbits, independent of values of the planets' masses (dark gray
region); one region that is stable (white region); and one region,
located between the other two regions, that is also unstable, but its
size depends on the masses of the planets. One example for the
  location of the green line, that mark the transition limit, will be
  given by equation (8). 

\begin{figure}[!t]
  \noindent
  \begin{tabular}{lr}
    \includegraphics[width=0.48\textwidth]{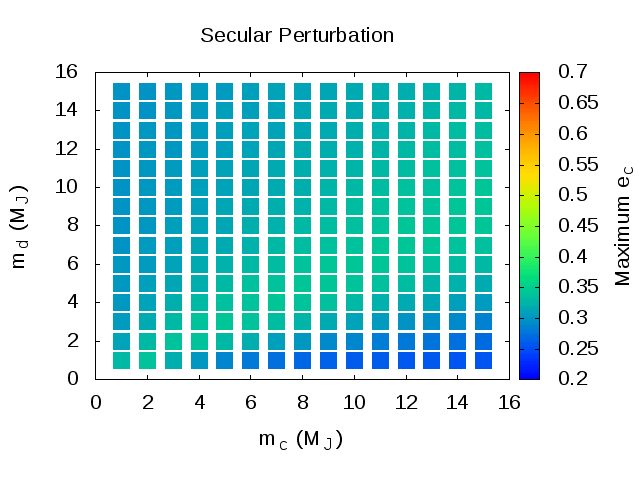}& 
    \includegraphics[width=0.48\textwidth]{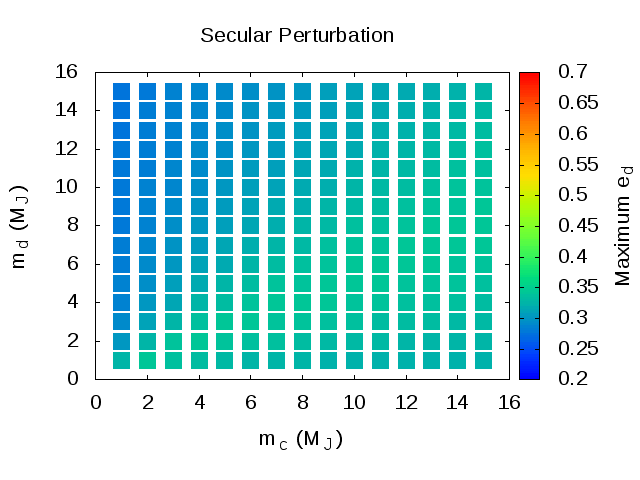}\\
  \end{tabular}
  \caption{The maximum values of eccentricity for each value of mass
    using secular perturbation. On the left plot are given the maximum
    values for the eccentricity of planet $c$ and, on the right plot
    for planet $d$. Both results presented eccentricity values below
    $0.35$, indicating only possibly stable orbits.}
  \label{fig:3}
\end{figure}

To evaluate the eccentricity evolution in different cases of
planets'masses in the system, we adopted a grid of masses ranging from
$1 M_J$ to $15 M_J$ with a step of $1 M_J$ for the masses of planets
$c$ and $d$.  We integrated for a total time of 30 thousand years.

We first analyze the maximum values of eccentricity for each value of
mass using the secular theory. We used the initial eccentricity values
of Curiel et al. (2011), given in Table 1.

Figures \ref{fig:3} shows the maximum values of eccentricity for planet $c$
(left plot) and for planet $d$ (right plot). The values in both cases
are all smaller than $0.5$. Considering that the initial values of the
eccentricities are approximately $0.26$ for planet $c$ and $0.30$ for
planet $d$, we see an eccentricity increase of less than $0.2$. So,
neither of them reach the critical eccentricity value. Consequently,
for the whole range of masses considered, the orbits do not cross each
other.

Therefore, according to the secular perturbation theory, all these
orbits are stable. However, the secular theory validity is limited due
to its approximations, which can be check through a comparison with
full numerical integrations.

\begin{figure}[!t]
  \noindent
  \begin{tabular}{lr}
    \includegraphics[width=0.48\textwidth]{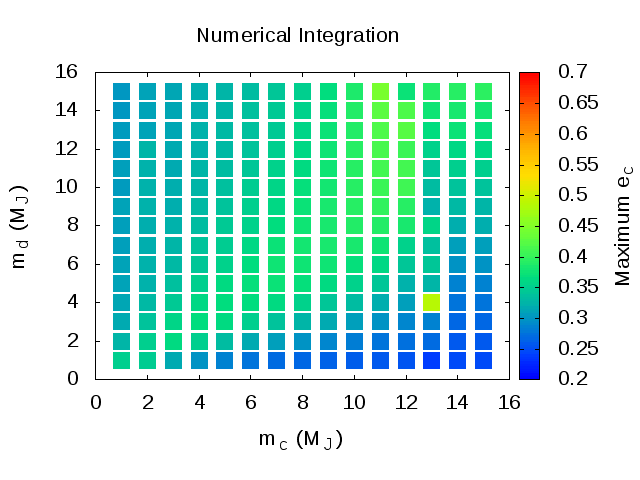}& 
    \includegraphics[width=0.48\textwidth]{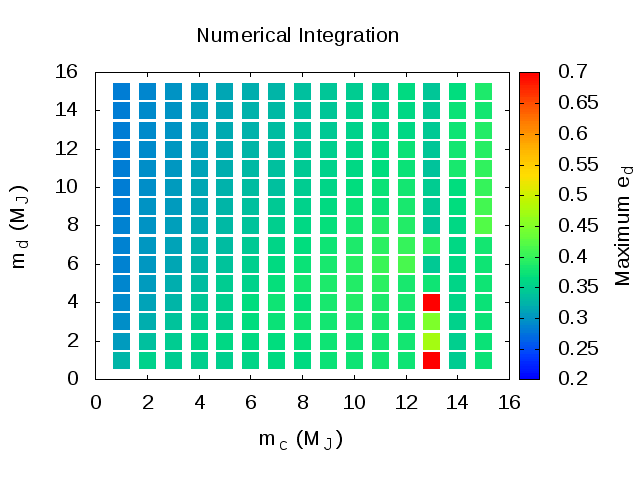}\\
  \end{tabular}
  \caption{The maximum values of eccentricity for each value of mass
    using numerical integration. On the left plot are given the
    maximum values of eccentricity for planet $c$ and, on the right
    plot for planet $d$. We can see larger values of eccentricities
    than in the secular theory case.}
  \label{fig:4}
\end{figure}

We proceed similarly using the {\it Mercury} package (Chambers, 1999)
to numerically integrate the system with different values of planetary
masses, throught the Burlish-Stoer integrator. When numerical
integrations are applied, short period terms are included which are
not accommodated by secular theory.

  In Figure \ref{fig:4}, the results for planet $c$ (left plot) do
  not present any case with an eccentricity larger than
  $0.5$. However, planet $d$ (right plot) has two possibilities of
  planets ejection, which occur at the red squares, meaning that the
  eccentricity is larger than the critical value and the orbits of the
  planets cross each other. The cases in the region of masses close to
  the ejection results (red squares) also present instability, but for
  the integration time used here (30 thousands years) there is no ejection,
  which does not exclude the possibility when integrated longer timescales.
  We performed a careful a visual inspection of the evolution of the 
  eccentricity  plots for all cases simulated in the grid of masses and we
  verified that from $m_c$ greater than $8 M_J$ the temporal evolution
   of the planets eccentricities show instability.

Comparing Figures \ref{fig:3} and \ref{fig:4}, we see that the secular
theory results reproduce the general trend of values of the maximum
eccentricities according to the planetary masses. However, these
values are smaller than the actual values, generated from the
numerical integrations.

In the present study, we aim to gauge the extent to which secular
theory can be used with the same accuracy as numerical integration.
To analyze the discrepancies between the two methods, we compare the
difference between the amplitudes of eccentricity variation for each
case of planetary mass, that is, the difference between the highest
value and the lowest value of the eccentricity of the planets.

\begin{figure}[!t]
  \noindent
  \begin{tabular}{lr}
    \includegraphics[width=0.48\textwidth]{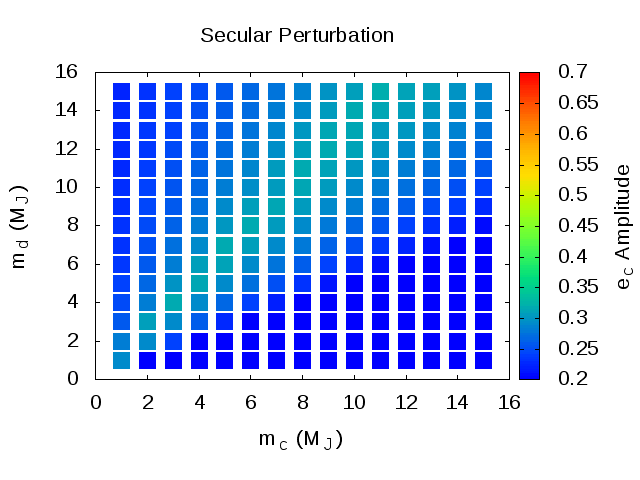}& 
    \includegraphics[width=0.48\textwidth]{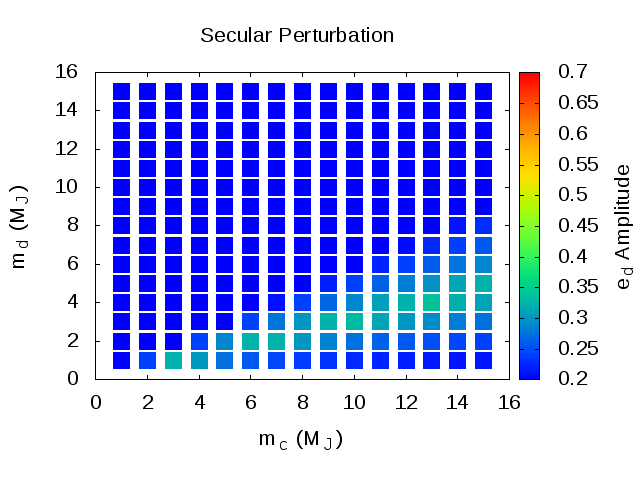}\\
  \end{tabular}
  \caption{Amplitude of the eccentricity variation for planets $c$ and
    $d$ according to the different values of masses (in Jupiter
    masses) using secular theory. On the color palette we have
    eccentricity amplitude values. On the left plot, we observe the
    amplitude of the eccentricity of planet $c$. The right plot shows
    the results for planet $d$.}
  \label{fig:5}
\end{figure}

  The plots in Figure \ref{fig:5} present the amplitudes of
  oscillation of the eccentricities of the planets for the case of the
  secular theory. These amplitudes are all smaller than $0.35$.

\begin{figure}[!t]
  \noindent
  \begin{tabular}{lr}
    \includegraphics[width=0.48\textwidth]{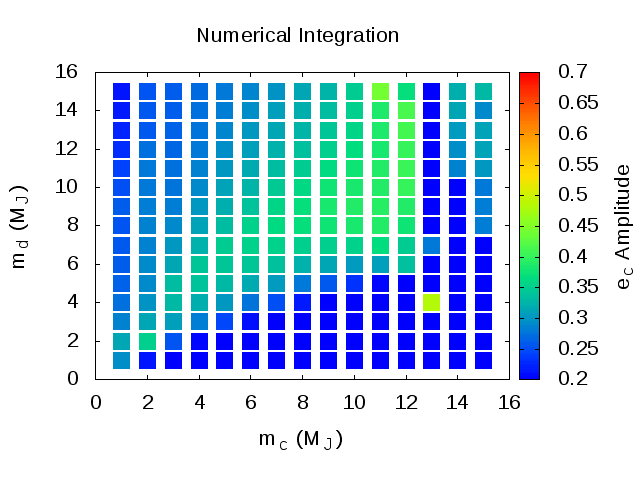}& 
    \includegraphics[width=0.48\textwidth]{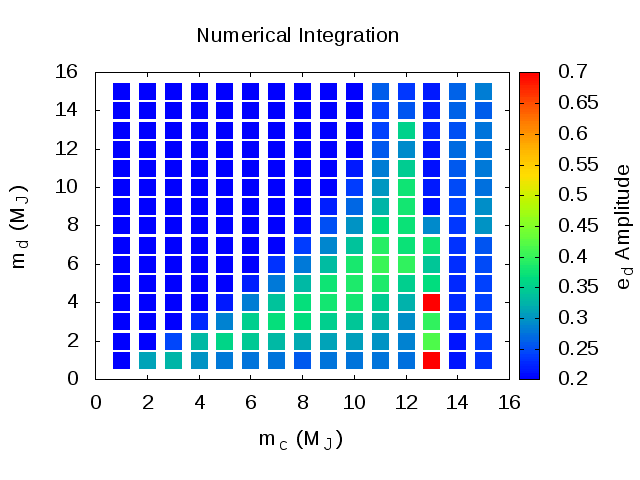}\\
  \end{tabular}
  \caption{Amplitude of the eccentricity variation for planets $c$ and
    $d$ according to the different values of masses (in Jupiter
    masses) for the numeral integration. On the color palette we have
    eccentricity amplitude. On left we observed the amplitude of the
    eccentricity of the planet $c$. On the right the figure shows the
    results for the planet $d$.}
  \label{fig:6}
\end{figure}

For the case of the numerical integrations, the amplitudes of
oscillation of the eccentricities are shown in Figure \ref{fig:6}.
For planet $c$ the amplitude can get up to $0.5$, while for planet $d$
there is a couple of orbits that reach values even higher.

\section{Validity for The Secular Theory}\label{sec:4}

In order to evaluate how good were the secular theory results in
comparison with the full numerical integration, we computed $\Delta
e_{num} - \Delta e_{sec}$, where $\Delta e_{num}$ corresponds to the
amplitude of oscillations of eccentricities from the numerical
integration and $\Delta e_{sec}$ is the amplitude of oscillation of
the eccentricities from the secular theory.

\begin{figure}[!t]
  \noindent
  \begin{tabular}{lr}
    \includegraphics[width=0.48\textwidth]{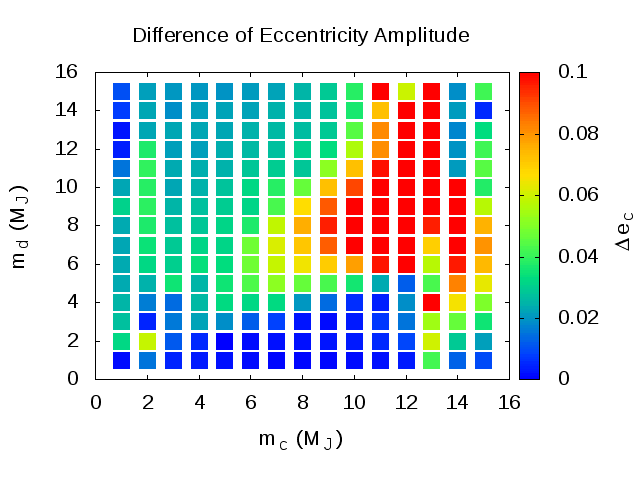}& 
    \includegraphics[width=0.48\textwidth]{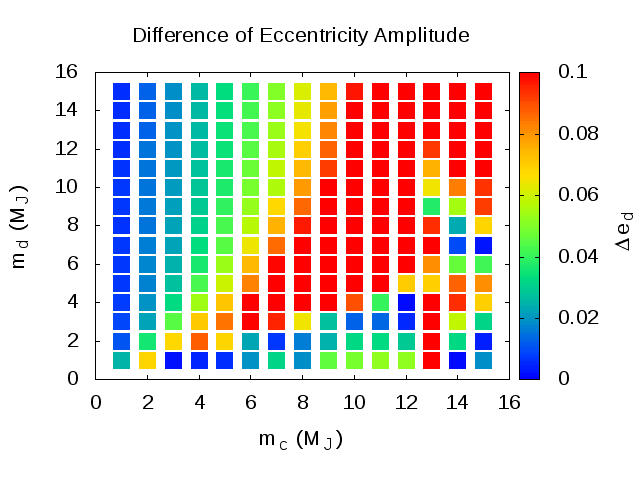}\\
  \end{tabular}
  \caption{ Difference of the eccentricities amplitude between the
    numerical integration results and the secular theory results for
    planets $c$( left) and $d$ (right) according to the variation of
    the masses (in Jupiter masses).}
  \label{fig:7}
\end{figure}

A first analysis of the results showed in Figure \ref{fig:7} indicates
that the secular theory works better for the inner planet (left plot)
than for the outer planet (right plot).  It is also better for lower
values of the masses. In the case of planet $c$ (inner planet) the
best results occurred for $m_{c}$ $\leq$ $7$$M_J$ and for $7$$M_J$ $<$
$m_{c}$ $<$ $12$$M_J$ with $m_{d}$ $\leq$ $5$$M_J$. In the case of
planet $d$ (outer planet) the best results occurred only for $m_{c}$
$<$ $4$$M_J$ with $m_{d}$ $>$ $2$$M_J$.

\begin{figure}[!t]
  \noindent
  \begin{tabular}{lr}
    \includegraphics[width=0.48\textwidth]{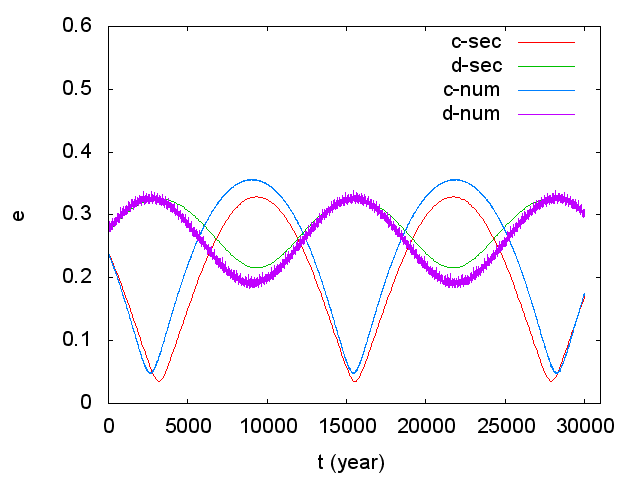}& 
    \includegraphics[width=0.48\textwidth]{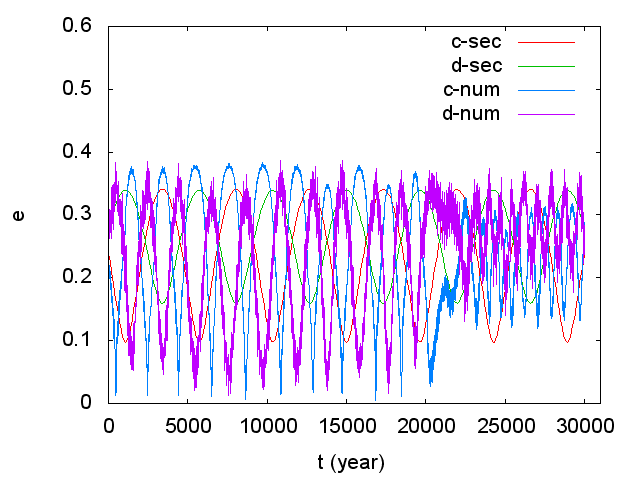}\\
  \end{tabular}
  \caption{ Evolution of eccentricity for planets $c$ and $d$.  The
    evolution of eccentricity by the secular theory is in red for
    planet $c$ and in green by the planet $d$. The eccentricity
    evolution by the numerical integration is in blue for planet $c$
    and in purple for planet $d$.  On the left plot, we have the
    stable case for $m_c = 3 M_J$ and $m_d$ $=$ $3$ $M_J$. In the
    right plot, we exemplify a chaotic case of eccentricity evolution,
    in this case $m_c$ $=$ $12$ $M_J$ and $m_d$ $=$ $8$ $M_J$.}
  \label{fig:8}
\end{figure}

In Figure \ref{fig:8}, we present two cases to exemplify the possible
destinations for the planetary dynamical evolution.  In the first
case, on the left plot, with planetary masses equal to $3$ $M_J$ for
planets $c$ and $d$, we have a stable case. This case could be
  used to draw the green line on figure 2, as we show in eqation (8) for
  the case $m_c = 13 M_J$ and $m_d = 1 M_J$. But, for this case,
  the green line will be located in a different position because the
  planetary masses are different and as a consequence the gravitational
  interactions change.

The planets eccentricity variation are very similar in numerical
integration and in secular theory.  Note that in the numerical
integration, planet $d$ has a secondary frequency. This secondary
frequency is due to the short period terms, which are considered in
numerical integration, but they are not included in the secular
theory. The second case, on the right plot, for the masses $m_c =
  12 M_J$ and $m_d = 8 M_J$ shows an unstable behavior for the
  numerical integration, but in the secular theory, as expected, we
  always have stable orbits.

\begin{figure}[!t]
  \noindent
  \begin{tabular}{lr}
    \includegraphics[width=0.48\textwidth]{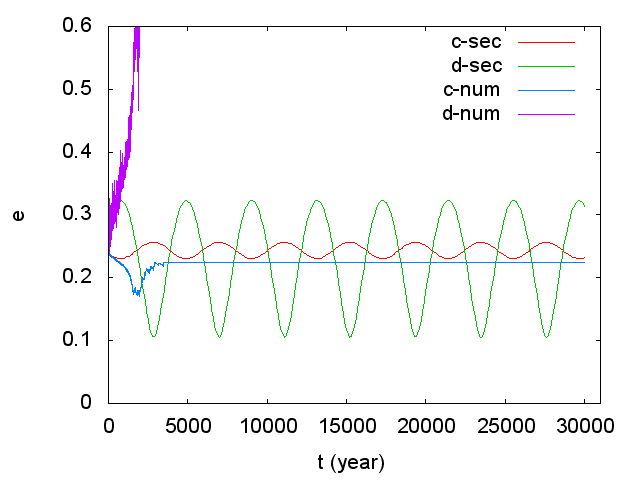}& 
    \includegraphics[width=0.48\textwidth]{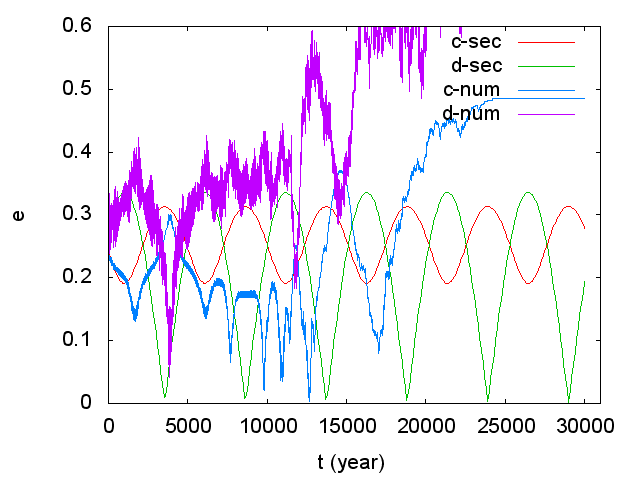}\\
  \end{tabular}
  \caption{Evolution of the eccentricity for planets c and d in the
    ejected cases.  The evolution of eccentricity in the secular
    theory is in red for planet $c$ and in green for planet $d$. The
    eccentricity evolution from the numerical integration is in blue
    for planet $c$ and in purple for planet $d$.  On the left plot, we
    have the case for the masses of $c$ equal to $13$ $M_J$ and $d$
    equal to $1$ $M_J$. In the right plot, we have the eccentricity
    evolution, for the masses of $c$ equal to $13$ $M_J$ and $d$ equal
    to $4$ $M_J$.}
  \label{fig:9}
\end{figure}

From Figure \ref{fig:4}, we have that when $m_c = 13 M_J$ and $m_d = 4
M_J$, both planets showed large variations in their eccentricities.
The orbit of the two planets were so unstable that planet $d$ was
ejected. The same occurred for the case when $m_c = 13 M_J$ and $m_d =
1 M_J$. The evolution of the eccentricity, in these two cases of
ejection, are shown in Figure \ref{fig:9}.  We consider that, in order
to ejection occurs in the numerical integration, the semi-major axis
has to be greater than $100$ $au$.

The left plot shows that planet $d$ is ejected in less than two
thousand years, while in the other case it is ejected much later, in
$25$ thousand years. In both cases the systems show a highly unstable
behavior since the beginning of the integration.

Therefore, the examples shown here cover the whole spectrum. A case of
planetary masses where the secular theory reproduces very well the
orbital evolution of the system (Figure \ref{fig:8}, left). A case
where the planetary masses are such that the evolution of the
eccentricities of the planets have their limits reproduced by the
secular theory, but not following the same pattern of of behavior
(Figure \ref{fig:8}, right). And finally, two cases where the
planetary masses are so big that the secular theory is of no use at
all (Figure \ref{fig:9}).

\begin{figure}[!t]
  \noindent
  \begin{tabular}{lr}
    \includegraphics[width=0.48\textwidth]{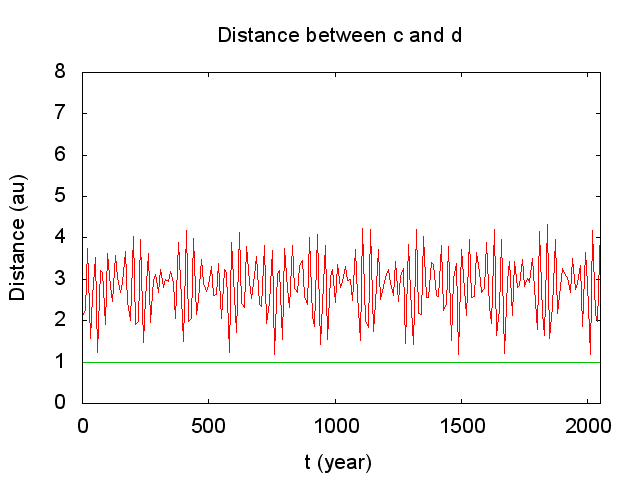}& 
    \includegraphics[width=0.48\textwidth]{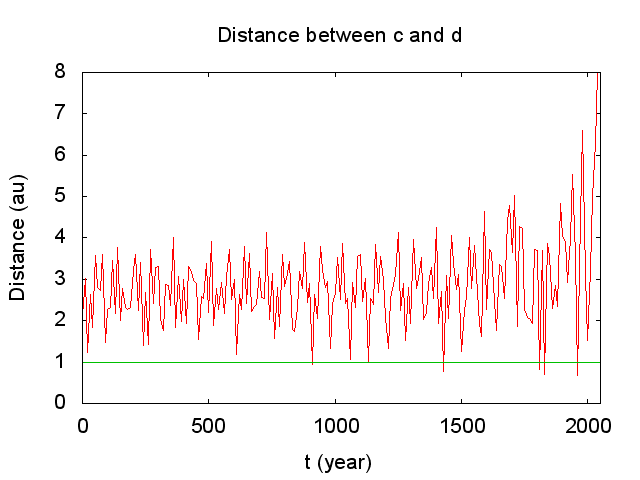}\\
  \end{tabular}
  \caption{ Distance between planets $c$ and $d$ for two thousand
    years.  On the left plot, we present the stable case, with masses
    of the two planets of $3$ $M_J$. On the right plot, the unstable
    case with ejection, for the masses of $c$ equal to $13$ $M_J$ and
    $d$ equal to $1$ $M_J$. The green line indicated the distance of
    $1$ $au$. }
  \label{fig:10}
\end{figure}

Next, we calculated the distance between the planets $c$ and $d$, to
try detect if there are close encounters between the two planets,
which could be the cause of the ejection of planet $d$. In Figure
\ref{fig:10}, we show the distance between the planets in a stable
case, on the left plot, with masses of the two planets equal to $3$
$M_J$.  We can observe that the planets are never less than one
astronomical unit from each other (green line).  Whereas, on the
right, we presented the ejection case, for the planets with mass $c$
equal to $13$ $M_J$ and $d$ equal to $1$ $M_J$, there is a close
encounter in less than one thousand years, the approach between the
planets decay to less than one astronomical unit.

Based on our discussion of close planetary encounters made in Figure
\ref{fig:2}, we have a limiting distance between the planets of 1~AU,
for the masses $m_c = 13 M_J$ and $m_d = 1 M_J$, so, equation (7) can be
rewritten as
\begin{equation}
  \begin{split}
    |a_c  (1 + e_c) - a_d (1 - e_d)| = 1 ,
  \end{split}
\end{equation}
where $a_c$ and $a_d$ are the semi-major axis of the planets $c$ and
$d$, respectively, and $e_c$ and $e_d$ are the eccentricities of the
planets $c$ and $d$.

Now, we can define a lower limit of eccentricities for planets with
such masses, beyond which the orbits would be unstable.  This limit
corresponds to the lower curve defining the Transition region shown in
Figure \ref{fig:2}.

All analyses reported in this study were conducted using a planar case
of a planetary system.  In cases with inclination it is possible that
planets with large masses have stable orbits and it can be explored in
future works.

\section{Conclusions}\label{sec:5}

The literature on the planetary system of $\upsilon$ Andromedae
presents a wide discrepancy for the planetary masses. In the present
work, we study the stability of planets $c$ and $d$ as a function of
their masses in order to contribute for the delimitation of their
possible masses.  We studied the temporal evolution of the
eccentricity of each planet using two approaches.  First adopting the
Lagrange-Laplace secular theory and then through the full numerical
integration.  We used the two methods to compare and check the limits
of the validity of the secular theory.

The results obtained in this work show that the use of secular theory
can infer the stability of a planetary system but only for a limited
range of values for the planetary masses.  For high values of mass,
the numerical integration becomes the best choice, mainly due the fact
that the secular theory does not take into account short period terms.

  With the numerical integration, we found the limits for the
  planetary masses to allow the orbital stability of the planets. For
  planetary masses larger than $8 M_J$ for planet $c$, independently 
  of the mass for planet d,   an unstable behavior is almost certain
  and possibilities of ejection of the planets exist.

As expected, we verified that above critical eccentricity values the
orbits cross each other, which leads to instability and possible
ejection. We also identified an unstable region without the need of
crossing orbits, whose size depends on the values of the
planets'masses.

\section*{acknowledgements}
  The authors are grateful for the support from CAPES, Fapesp - proc
  2016/24561-0 and CNPq - proc 312813/2013-9. We thank Gabriel
  Borderes Motta and Alexandros Ziampras for the suggestions. We also
  would like to acknowledge the hard work made by the referees.

\end{document}